\documentclass[british]{article}
\usepackage[T1]{fontenc}
\usepackage[latin9]{inputenc}
\usepackage{listings}
\usepackage{tipa}
\usepackage{amsmath}
\usepackage{amssymb}
\usepackage{mathdots}
\usepackage{graphicx}
\usepackage{esint}
\PassOptionsToPackage{version=3}{mhchem}
\usepackage{mhchem}

\makeatletter

\providecommand{\tabularnewline}{\\}

\makeatother

\usepackage{babel}
\begin{document}

\title{A new kind of complexity}

\author{Rade Vuckovac\date{}}

\maketitle

\begin{abstract}
A new class of functions is presented. The structure of the algorithm,
particularly the selection criteria (branching), is used to define
the fundamental property of the new class. The most interesting property
of the new functions is that instances are easy to compute but if
input to the function is vague the description of a function is exponentially
complex. This property puts a new light on randomness especially on
the random oracle model with a couple of practical examples of random
oracle implementation. Consequently, there is a new interesting viewpoint
on computational complexity in general.
\end{abstract}

\section{Introduction\label{sec:Introduction}}

The structured program theorem, also known as Böhm-Jacopini theorem
is one of the premises for this paper. The theorem shows an ability
of the algorithm or a program implementing that algorithm to compute
any computable function combining only the three algorithmic structures.
These structures are:
\begin{itemize}
\item The tasks in the program are done one after another one (sequential
order of execution)
\item The program can branch to a different path of execution depending
on some statement evaluation result (selection criteria)
\item Repeating some task until an evaluation of some statement is satisfied
(iteration)
\end{itemize}
The branching or selection criteria structure and its usage defines
the new proposed class of functions. It is assumed that structured
program theorem holds and that usage of the three structures is Turing
complete. More precisely, it is assumed that further reduction from
three structures to two structures (excluding branching) is impossible.
In other words the selection criteria cannot be effectively replaced
with a combination of the other two structures.

The second premise is the analysis of the branching structure in software
metrics done by McCabe \cite{key-1}. The main result of McCabe's
work is the notion of Cyclomatic Complexity (CC). The flow chart of
the above mentioned structures is used to count the individual execution
paths that the program can take. The CC is mainly used as software
testing metric. It evaluates a requirement of how many testing cases
are needed for a piece of software. In the majority of cases the relation
between branching and individual paths is exponential, meaning that
if the number of branching in a program increases then the number
of individual paths which the program can execute doubles for every
added branching. 

The combination of the two above notions can lead to an extraordinary
case. The program can be written with a non restricted number of branching
($n$) implying an exponential growth of the number of paths that
the program can take through execution ($2^{n}$). Additionally, the
theory of algorithms demands that the formal description of an algorithm
shall include every possible case it can take through execution.
\begin{quote}
``Typically, when an algorithm is associated with processing information,
data is read from an input source, written to an output device, and/or
stored for further processing. Stored data is regarded as part of
the internal state of the entity performing the algorithm. In practice,
the state is stored in one or more data structures.

For some such computational process, the algorithm must be rigorously
defined: specified in the way it applies in all possible circumstances
that could arise. \emph{That is, any conditional steps must be systematically
dealt with, case-by-case; the criteria for each case must be clear
(and computable)} {[}emphasis added{]}.

Because an algorithm is a precise list of precise steps, the order
of computation will always be critical to the functioning of the algorithm.
Instructions are usually assumed to be listed explicitly, and are
described as starting \textquotedbl{}from the top\textquotedbl{} and
going \textquotedbl{}down to the bottom\textquotedbl{}, an idea that
is described more formally by flow of control.'' \cite{key-3}
\end{quote}
That means that an algorithms with a high CC cannot be practically
described because the number of execution paths increases exponentially.
On the other hand, the instances of such an algorithm can be easily
computed because the increase of the number of branching in the program
incurs only polynomial cost.

This extraordinary case needs more thorough clarifications:
\begin{itemize}
\item It is not clear what the irreducible number of branching means. There
is still a possibility that individual execution paths of an algorithm
are actually identical transformations.
\item Although CC shows exponential dependency between the number of branching
and the number of execution paths, a possibility remains that the
relationship between the selection criteria and the paths doubling
numbers can be reduced to an acceptable level. Indeed, there is a
suggestion to avoid high CC software metric: the first is to rewrite
a program in question with reduced use of branching, and the second
is to split the program in more manageable pieces \cite{key-4}. 
\end{itemize}
These concerns and other relevant discussions are explored through
3n+1 problem (section \ref{sec:Deconstruction-of-the}) and Wolfram's
rule 30 (section \ref{sec:Wolfram's-rule-30}). Section \ref{sec:Conclusion}
summarises the new function features and speculates on the impacts
on randomness and P/NP classes.

\section{Programming take on 3n+1 problem\label{sec:Deconstruction-of-the} }

\subsection{3n+1 problem }

The 3n+1 problem is ideal for exploring the relationship between the
algorithm's selection criteria and the CC. One reason for this is
that the 3n+1 problem is extensively studied and a lot of details
about the problem are well established. Another reason is that the
selection criteria are essential part of the problem description.
The problem is very simple to state: take any positive integer, if
the integer is even divide that integer with 2, if the integer is
odd multiply the integer with 3 and add 1. Repeat the procedure until
the result is 1. The problem is to decide if all positive integers
reach 1. One step of the 3n+1 problem is shown below in modular notation
\cite{key-6}.

$f(n)=\begin{cases}
n/2 & \text{if \ensuremath{n\equiv0\text{ (mod 2)}}}\\
3n+1 & \text{if \ensuremath{n\equiv1\text{ (mod 2)}}}
\end{cases}$

It is possible to skip the evaluation if the integer is even after
$3n+1$ operation and proceed with the operation $n/2$ because 3
times odd integer plus one is even. That is also an example of how
CC of an algorithm can be reduced. Below is the optimised version
which will be used throughout the rest of the paper. 

$f(n)=\begin{cases}
n/2 & \text{if \ensuremath{n\equiv0\text{ (mod 2)}}}\\
(3n+1)/2 & \text{if \ensuremath{n\equiv1\text{ (mod 2)}}}
\end{cases}$

\begin{figure}[h]
\includegraphics[scale=0.5]{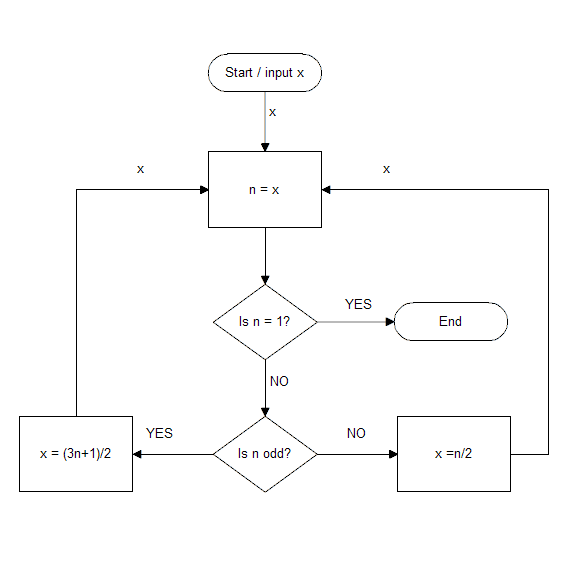}

\caption{\label{3n+1 chart}3n+1 flow chart with optimised (3n+1)/2 step}
\end{figure}

The 3n+1 problem flow chart is shown in figure \ref{3n+1 chart}. 

\subsection{3n+1 and Cyclomatic Complexity}

\begin{figure}[h]
\includegraphics[scale=0.5]{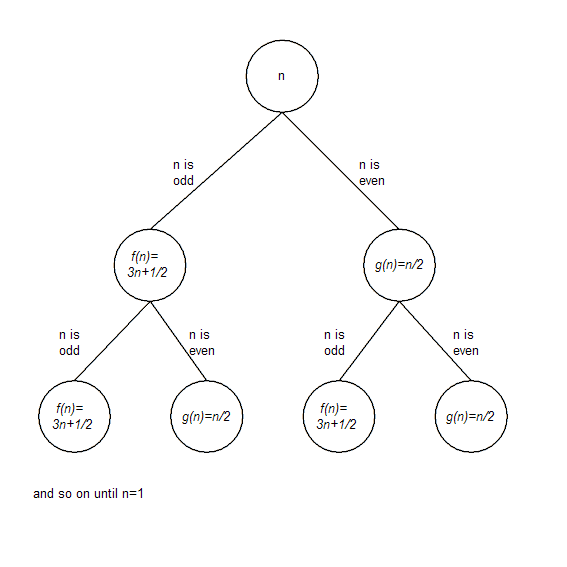}

\caption{\label{3n+1 tree}two steps in 3n+1 algorithm (binary tree resemblance)}

\end{figure}

In this section the CC of 3n+1 algorithm is explored. As mentioned
in section \ref{sec:Introduction} CC is software metric. That software
metric measures how many paths the program can take through the execution.
It uses graph theory to count individual execution paths. The formula
for CC is as follows: $\nu(G)=e-n+2$ where $e$ and $n$ are amounts
of edges and nodes contained in the algorithm flow graph. $\nu$ is
cyclomatic number and $G$ means that the complexity is the function
of algorithm flow graph \cite{key-4}. Applying this formula to a
particular algorithm is a not straight forward exercise. One of simpler
ways is counting the binary decision predicates $p$ . The formula
for this approach is $\nu(G)=p+1$. The figure \ref{3n+1 tree} shows
the 3n+1 algorithm doing two steps. Using a simpler method the three
binary branching can be identified, therefore $\nu(G)=3+1$ and indeed
there are four individual paths the 3n+1 algorithm can take in two
steps. It is evident from figure \ref{3n+1 tree} that every 3n+1
step will double CC. That means after doing several 3x+1 steps the
system starts to be very complex from the software testing perspective.

\subsection{3n+1 Preliminaries\label{sub:3n+1-Preliminaries}}

A few details about 3n+1 problem are well known and mentioned here
\cite{key-6}, some are listed for further discussion:
\begin{itemize}
\item The experimental data confirms that numbers up to $\approx2^{60}$are
reaching one\cite{key-8}.
\item The lower bound of how many natural numbers reach one is shown by
\cite{key-5}. For any sequence of natural numbers in the interval
$[1.x]$ the number of naturals reaching one in corresponding interval
is $>x^{0.84}$.
\item A parity sequence for each natural number as an input is unique and
that is true even if not all natural numbers are reaching one. The
parity sequence is formed by putting 1 or 0 in the sequence, depending
on what operation is performed in the particular step. In other words
if the branching in figure \ref{3n+1 chart} ``is n odd?'' is NO
put $0$ in the parity sequence and if it is YES put $1$.
\end{itemize}

\subsection{3n+1 as composite function}

The composite function nature of 3n+1 problem comes from the parity
sequence. For example if $n=13$ the parity sequence for that input
will be $1,0,0,1,0,0,0$. In the same manner a composite function
$h(n)$ can be composed, for example $h(13)=f\circ g\circ g\circ f\circ g\circ g\circ g=1$
where $f(n)=(3n+1)/2$ and $g(n)=n/2$. It is obvious that the parity
sequence pattern and the composite function pattern are identical.
That should be expected because the parity sequence is the description
of how a natural number is transformed under the 3n+1 rule.

If the natural numbers and their corresponding parity sequences have
the bijective relationship, and that appears to be true even the $1$
is not periodicity revolving point \cite{key-6}, then the natural
numbers and their corresponding composite function are bijective as
well. That is based on the fact that the pattern of the parity sequence
is identical to the pattern of the composite function for the same
natural number.\emph{ Therefore every natural number has a unique
composite function $h(n)$ to map a natural number to a number where
a period occurs under rule }3n+1\emph{.}

\subsection{\label{sub:3n+1-as-encoding system}3n+1 as encoding system}

A parity sequence or $h(n)$ pattern can be considered as binary encoding
for every natural number reaching 1 under 3n+1 rule. For example for
the number $13$ its binary representation is $1101$. In 3n+1 binary
world encoding for $13$ is FG string $fggfggg$. The decoding is
done by applying the rule 3n+1 backwards without the need for evaluating
``if odd or even''. Just start from one and read the FG string backwards.
If the character is $g$ apply function $g(n)=2n$ and go to next
character. If the character is $f$ apply function $f(n)=(2n-1)/3$
and go to the next character. When all characters are read the final
number is the decoded number. From traditional binary and parity encodings
two languages can be defined:
\begin{enumerate}
\item The binary language $L_{0,1}$ is written by $\{0,1\}$ alphabet.
The members of $L_{0,1}$ set are the binary encodings of all natural
numbers reaching 1 under rule 3n+1 $(n')$. The ratio size of the
set is $\geqq n^{0.84}$\cite{key-5}.
\item The parity language $L_{f,g}$ is written by $\{f,g\}$ alphabet.
The members of $L_{f,g}$ set are the parity sequences of $n'$. The
size of $L_{f,g}$ is the same as the size of $L_{0,1}$.
\end{enumerate}
Although the amounts of the words forming languages $L_{0,1}$ and
$L_{f,g}$ are equal the words explaining the same object differ in
length between languages. For example encodings for number 13 are:
\begin{enumerate}
\item In $L_{0,1}$language the encoding is $1101$; size of the word is
4. 
\item In $L_{f,g}$language the encoding is $fggfggg$; size of the word
is 7. 
\end{enumerate}
It is obvious that encodings from language $L_{f,g}$ can be compressed,
but compression can not go below complexity of the language $L_{0,1}$.
This means that \emph{the complexity of language }$L_{f,g}$\emph{
(}3n+1\emph{ encoding) is greater or equal to complexity of language
}$L_{0,1}$\emph{ (optimal binary encoding).}

A significant implication is that the 3n+1 function description depends
on input in an unusual way. The input is not just an ordinary variable
but it is determinant of how a particular transformation (from input
towards one) is composed. If the if else structure is used in 3n+1
rule then the composite function description for all natural numbers
has at least sub exponential growth of $n^{0.84}$.%
\footnote{Note that first $\approx2^{58}$natural numbers are reaching one and
even if the some numbers do not reach one the corresponding parities
are unique. %
}

\subsection{3n+1 as a random function?\label{sub:3n+1-as-a oracle}}

Random oracle is an abstraction used to model security protocols and
schemes. Basically random oracle is an imaginary machine which upon
an input to oracle, randomly draws a function from a set of all function
possible and with that function an output is calculated and returned.
A simple model can be used as an example: On input $0$ flip fair
coin and record the resulting tail/head occurrences as a truly random
binary string; continue with same procedure for inputs $1,00,01,10,11,000\ldots$
(see table \ref{tab:Random-Oracle-mapping}).

\begin{table}[h]
\begin{tabular}{|c|c|}
\hline 
binary input & corresponding string \tabularnewline
\hline 
\hline 
$0$ & truly random string 1\tabularnewline
\hline 
$1$ & truly random string 2\tabularnewline
\hline 
$00$ & truly random string 3\tabularnewline
\hline 
$01$ & truly random string 4\tabularnewline
\hline 
$\ldots$ & $\ldots$\tabularnewline
\hline 
\end{tabular}

\caption{\label{tab:Random-Oracle-mapping}Mapping using random function}

\end{table}

The table \ref{tab:Random-Oracle-mapping} is then used in proving
various security systems (see \cite{key-10} for details). It is apparent
that the table \ref{tab:Random-Oracle-mapping} is not practical by
means of storage and access to intended entry. In practice random
oracle is replaced with cryptographically secure hash with undefined
security consequences. The work of \cite{key-11} argues that random
oracle modelling is essentially unsound; a practical implementation
of replacing a random oracle in proven secure scheme results in an
insecure scheme. 

An interesting property defined in \cite{key-11} is a notion of correlation
intractability. The correlation intractability is the resistance to
put some relation between inputs and outputs on some mapping. It is
easy to see that random oracle is resistant to correlation (table
\ref{tab:Random-Oracle-mapping}) because of flipping fair coin. For
potential replacement, and that is single functions or function assemblies,
correlation intractability property can not be guaranteed. The reasoning
behind is that mapping description is shorter than allowed input description
used by adversary, therefore the correlation between input and output
must exist and that can not be expected from efficient and fully described
function or function assembly to behave randomly. Quotes from \cite{key-11}
\begin{quote}
Informal Theorem 1.1 There exist no correlation intractable function
ensembles... The proof of the above negative result relies on the
fact that the description of the function is shorter than the input
used in the attack.

Correlation Intractability. In this section we present and discuss
the difficulty of defining the intuitive requirement that a function
ensemble behaves like a random oracle even when its description is
given. We first comment that an obvious over-reaching definition,
which amounts to adopting the pseudo-random requirement of {[}12{]},
fails poorly. That is, we cannot require that an (efficient) algorithm
that is given the description of the function cannot distinguish its
input-output behaviour from the one of a random function, because
the function description determines its input-output behaviour.
\end{quote}
Despite that 3n+1 shall apply for random oracle replacement. One line
of argument can go along the fact that 3n+1 is perceived as a hard
problem. Quotes from \cite{key-13}p4 and p17:
\begin{quote}
The track record on the 3 x + 1 problem so far suggests that this
is an extraordinarily difficult problem, completely out of reach of
present day mathematics. Here we will only say that part of the difficulty
appears to reside in an inability to analyze the pseudorandom nature
of successive iterates of T ( x ), which could conceivably encode
very difficult computational problems.

The iterates of the shift function are completely unpredictable in
the ergodic theory sense. Given a random starting point, predicting
the parity of the n-th iterate for any n is a \textquotedblleft{}coin
flip\textquotedblright{} random variable. 
\end{quote}
One obvious advantage of replacing table \ref{tab:Random-Oracle-mapping}with
table \ref{tab:3n+1-parity-mapping}is that entries in 3n+1 parity
table can be produced \emph{deterministically}. Finding any pattern
or structure in table \ref{tab:3n+1-parity-mapping} may open a way
to attack the 3n+1 problem. A similar argument is made with hardness
of integer factoring and consequent factoring use in asymmetric encryption.

The second line of replacing random oracle with parity sequences is
complexity of the 3n+1 in terms of CC and composite function model.
If 3n+1 is considered as composite function, the form without specifying
input looks like formula \ref{eq:composite uni} where $f\clubsuit g$
means depending on input use function $f$ or $g$. That can not be
considered as \emph{a fully described function}. Only with an input
the formula can make sense (and can be executed).

\begin{equation}
(f\clubsuit g)\circ(f\clubsuit g)\circ(f\clubsuit g)\ldots\label{eq:composite uni}
\end{equation}

The argumentation can also go along the line input and function description
equality. As is shown in subsection \ref{sub:3n+1-as-encoding system}
input language and composite function (parity) language for 3n+1 are
of equal complexity. The configuration where input description and
function description are of the same length, is actually listed as
a possible case where random function can be replaced (see restricted
correlation intractability section \cite{key-11}). Although that
case is considered as inefficient, as is table \ref{tab:Random-Oracle-mapping}
for example (function description is actually input/output description).
However table \ref{tab:3n+1-parity-mapping} is practical because
entries can be calculated as is needed (full knowledge of all mappings
are not necessary).

Here is how the 3n+1 implementation of the hash function (random function
replacement) may look: Let the input $n$ be a word with at least
$256$ bits in length. Treat $n$ as an unsigned integer. Process
$n$ by the algorithm figure\ref{3n+1 chart}. Form the binary sequence
(parity) by recording $1$ when ``yes'' and $0$ when ``no'' is
answered to the question ``is n odd?''. Stop when parity is $128$
bit long. The game is to find $n'$ in the way to produce identical
first $128$ bits in parity sequence as $n$ does (a collision). The
search for collision is needed for a specific input, because the powers
of two ($32,64,128\ldots$) inputs will produce parities of zeros
(collisions are trivial, see entry $8$ in table \ref{tab:3n+1-parity-mapping}
for example). Because there is only formula \ref{eq:composite uni}
and target parity for someone who wants to find the match for that
parity, the task is impossible excluding exhaustive search. 

\begin{table}
\begin{tabular}{|c|c|}
\hline 
$n$ & corresponding parity \tabularnewline
\hline 
\hline 
$\ldots$ & $\ldots$\tabularnewline
\hline 
$7$ & $111010001000$\tabularnewline
\hline 
$8$ & $000$\tabularnewline
\hline 
$9$ & $10111010001000$\tabularnewline
\hline 
$\ldots$ & $\ldots$\tabularnewline
\hline 
\end{tabular}

\caption{\label{tab:3n+1-parity-mapping}3n+1 parity mapping}

\end{table}

\subsection{3n+1 inherently serial problem?\label{sub:3n+1-inherently-serial}}

An Inherently serial problem is when some algorithms can not be split
into chunks and executed concurrently to reduce execution time. The
reason is that the next execution step in such an algorithm depends
on previous step result \cite{key-19}. A formal attempt to define
sequential nature of algorithm with example can be found here \cite{key-18}.
Because 3n+1 algorithm is essentially a composite function, it is
evident that the 3n+1 algorithm step fundamentally depends on the
result from previous step, see figures \ref{3n+1 tree} and \ref{3n+1 chart}.
That means that the 3n+1 algorithm as is formulated in figure \ref{3n+1 chart}
can not be divided and executed in parallel and be more efficient
than serial execution.

Computational irreducibility (CI) notion introduced by Wolfram \cite{key-23}is
closely related to serial phenomena in algorithms. CI in the cellular
automata (CA) world means that the fastest way to have knowledge of
what a particular CA is doing is to run that CA. The same observation
can be applied to the 3n+1 algorithm. CI is more obvious in 3n+1 than
in other systems because 3n+1 is a composite function with consequence
of inherent non parallelism.

\subsection{3n+1 and reductions\label{sub:3n+1-and-reductions}}

\begin{figure}[!h]
\includegraphics[scale=0.5]{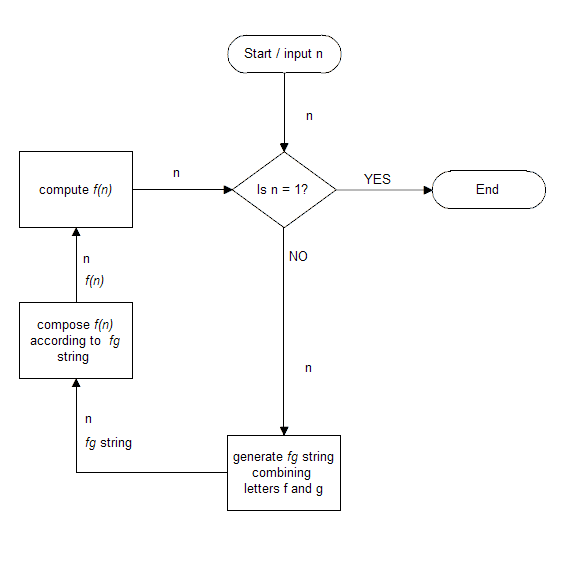}

\caption{\label{3n+1 inefficient}3n+1 flow chart with Do-While structure (looping)}

\end{figure}

It was tried before to show that the 3n+1 problem is intractable.
One example is here \cite{key-9}. The main argument of that work
goes on showing that the 3n+1 solution has to be infinitely complex,
using Solomon-Kolgomorov-Chaitin (SKC) complexity as an argument\cite{key-14}.
It relies on the fact that every 3n+1 transformation is unique and
if we were to represent all of them, the only remaining option would
be to list them all and consequently that option is obviously infeasible.
It is similar reasoning to the one from section \ref{sub:3n+1-as-encoding system}.
The problem with either reasoning is the possibility that 3n+1 inquiries
might be calculated by some algorithm other than algorithm shown in
figure \ref{3n+1 chart} and furthermore that the other algorithm
can be fundamentally different. It is impossible to know how that
algorithm may look anyhow a couple of important properties can be
defined:
\begin{itemize}
\item \emph{low CC}; Only algorithms without using branching structure can
be considered as candidates. 
\item \emph{efficiency}; There are algorithms with low CC see figure \ref{3n+1 inefficient}
for example. The execution time of that algorithm depends on the oracle
proposing the $fg$ string (as shown in subsection \ref{sub:3n+1-as-encoding system}).
One option for getting the answer from that algorithm is that the
oracle goes through an exhaustive search to match $fg$ string with
output $1$ (if 3n+1 conjecture holds). To be efficient it is required
from the candidate algorithm to produce a matching $fg$ string by
evaluating input $n$ in P time.
\end{itemize}
If both above requirements are met by the candidate algorithm, then
the apparent CC of 3n+1 can be reduced in P time. The algorithm on
presented input $n$ can predict the branching $fg$ string without
using branching structure. Consequently selection programming structure
can be replaced by a combination of sequence and iteration without
significant cost (P time). In that case the structural programming
theorem \cite{key-2} needs revision.

\section{About Wolfram's rule 30\label{sec:Wolfram's-rule-30}}

\subsection{Rule 30 complexity\label{sub:Rule-30-complexity}}

The Rule 30 is probably one of the most represented rules in the Wolfram's
NKS book \cite{key-20}. The definitions of rule 30 are listed below:
\begin{itemize}
\item Boolean form is \cite{key-16} $p$ $Xor$ $(q$ $Or$ $r)$
\item English description \cite{key-17} : \end{itemize}
\begin{quote}
``Look at each cell and its right-hand neighbor. If both of these
where white on the previous step, then take the new color of the cell
to be whatever the previous color of its left-hand neighbor was. Otherwise,
take the new color to be opposite of that''.\end{quote}
\begin{itemize}
\item Visual description and example is shown in figure \ref{rule 30}.
\end{itemize}
The main features of the Rule 30 are chaotic behaviour and randomness.
Both features are accomplished by an apparently simple rule and with
an input with only one black cell - see figure \ref{rule 30}. Quote
from NKS book \cite{key-20} pages 27-28. 
\begin{quote}
``The picture shows what happens when one starts with just one black
cell and then applies this rule over and over again. And what one
sees is something quite startling--and probably the single most surprising
scientific discovery I have ever made. Rather than getting a simple
regular pattern as we might expect, the cellular automaton instead
produces a pattern that seems extremely irregular and complex. But
where does this complexity come from? We certainly did not put it
into the system in any direct way when we set it up. For we just used
a simple cellular automaton rule, and just started from a simple initial
condition containing a single black cell.''
\end{quote}
That observation is mentioned numerous times and is not entirely correct
on both accounts (simple rule, one black cell as input). Let us use
figure \ref{rule 30} for example. 
\begin{itemize}
\item The first row shows the input and it is \emph{43 bits long} with 42
white cells and one black. Instead talking of only one black cell
input, emphasis should be on low entropy of that input. Also it should
be explained how entropy of the input is relevant to the rule 30 process,
because the configuration with one black cell has the same probability
of occurring as any other configuration.
\item The English description of the rule already mentioned is actually
the clue to chaotic / random behaviour. The description is as follows:
if something is true do that else do something different. It is exactly
the same structure already seen in the 3n+1 problem. Considering that,
rule 30 can be considered as composite function in the same fashion
as 3n+1. The difference between 3n+1 and rule 30 is that the rule
30 update of cell depends on outputs of neighbouring cells as well.
Therefore it is trickier to calculate CC of rule 30 algorithm. A short-cut
to estimating CC is to assume one branching per row evolution. Since
rule 30 (figure \ref{rule 30}) is iterated 21 times, the amount of
possible execution paths for one cell is $2^{21}$. From the software
testing perspective anything over $2^{10}$ is practically non testable
\cite{key-4}.
\end{itemize}
Consequently, it is not correct to brand rule 30 as a simple program
while at the same time it has an inherently high level of CC.

\subsection{Rule 30 function description\label{sub:Rule-30-function description}}

Although the rule 30 algorithm is fairly simple, its function description
is certainly complex. The reason for this is that a particular input
and particular number of iterations actually define which composite
function is going to be executed at the time. Unlike the 3n+1 case
where input alone determines the number of iterations and consequently
CC, the rule 30 CC depends on input and the number of iterations.
Quantifying rule 30 CC is shown below.

\begin{figure}[h]
\includegraphics[scale=0.33]{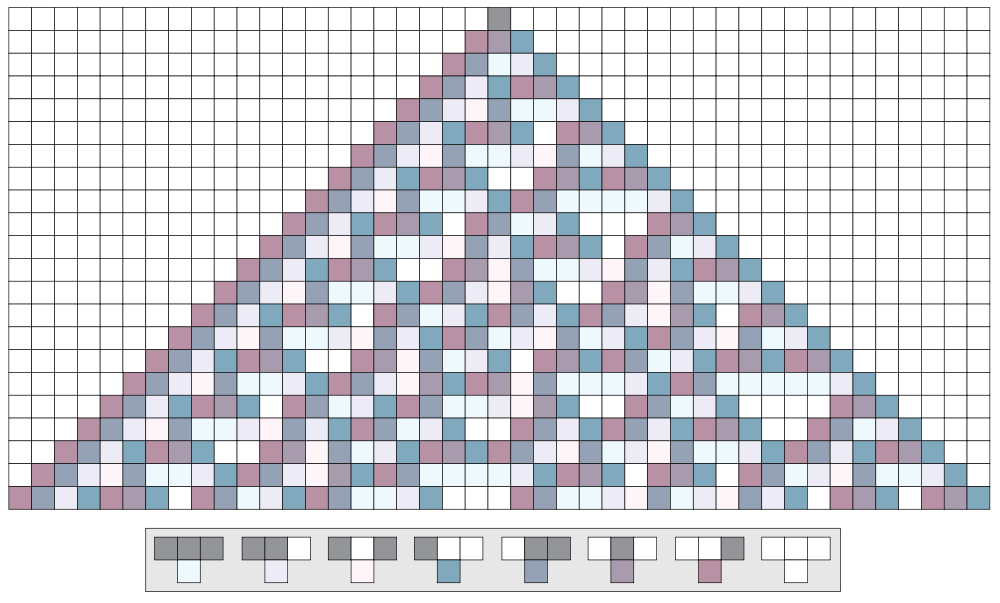}

\caption{\label{rule 30}rule 30, example of evolution and rules of transformation,
image copied from \cite{key-15}}

\end{figure}

Let $ln$ be the length of the input ($ln=43$ figure \ref{rule 30})
and $li$ be the number of iterations ($li=21$figure \ref{rule 30}).
If $li\leq ln$ CC depends on $li$ , and the number of paths $\nu(G)$
that the algorithm can take is $\nu(G)=2^{li}$. In the case $li>ln$
the number of paths is $\nu(G)=2^{ln}$. The reason for that is that
the entropy of the number of paths is bounded by entropy of input.
Therefore the number of paths that the rule 30 algorithm can take
through execution is:

$\nu(G)=\begin{cases}
\nu(G)=2^{li} & \text{if \ensuremath{li>ln}}\\
\nu(G)=2^{ln} & \text{if \ensuremath{ln>li}}
\end{cases}$

This means that if input length is smaller than number of iteration,
CC depends on input only as it is the case with 3n+1. Wolfram in his
NKS \cite{key-20} uses empirical methods to argue on some rule 30
attributes. For example empirical data shows that the period of rule
30 has an exponential growth in relation to input, which indicates
that above assertion of exponential growth is true.

\subsection{Rule 30 as hash function}

Having the same algorithm structure as 3n+1, rule 30 is also a candidate
for the hash function. There is a proposal which appeared on sci.crypt
\cite{key-21}
\begin{quote}
``Let length of constant c be a desired length of a hash. Constant
c can be arbitrary chosen. For example if 128 bits hash is required
the constant c may easily be 128 zeros. The string s for hashing is
then concatenated to the constant c to form a starting row r for rule
30; r = c + s. The row r is then evolved twice row length. For example
if c = 128 bits in length and s = 128 bits in length then evolution
is performed 512 times (column length is 512). Now the part (length
of c) of last row serves as a hash. From above example the first 128
bits of 512th row is the hash h of the string s.'' 
\end{quote}
The two major points raised in the discussion are the efficiency of
algorithm and the choice of $c$ to be string of zeros. Even though
the proposed hash is not practical (quadratic in nature) it is still
in $P$. Constant $c$ instead of zeros should employ some pseudo-random
string such as $\pi$ number sequence to avoid short cycle of rule
30.

\subsection{Rule 30 reduction\label{sub:Rule-30-and irreducibility}}

Rule 30 apparently satisfies two requirements discussed in subsection
\ref{sub:3n+1-and-reductions}: low CC and efficiency. Both requirements
for rule 30 are apparently satisfied by use of boolean form instead
English form (subsection \ref{sub:Rule-30-complexity}). 

Determining branching structure presence in CA is not an easy task.
For example, Wolfram rule 110 is Turing complete \cite{key-25} and
it has to accommodate use of if/else structure somehow (the structured
programming theorem \cite{key-2}). On the other hand it is not entirely
clear how the 110 compiler can be employed for usual programming tasks
including branching \cite{key-7}. However the branching structure
of rule 110 appears in English description \cite{key-25} as well.
\begin{quote}
``The values are updated in the following way: 0s are changed to
1s at all positions where the value to the right is a 1, while 1s
are changed to 0s at all positions where the values to the left and
right are both 1.''
\end{quote}
It is assumed that the boolean form and the English form of rule 30
are equal when quantifying CC. Another argument for that is: a tracing
of rule 30 program execution paths produces the same binary tree structure
as CC structure in figure \ref{3n+1 tree}. Which path is going to
be executed depends entirely on the rule 30 input. Steps through rule
30 evolution can not be skipped. Even though the branching structure
is not explicitly present in rule 30 boolean implementation, CC structure
is present.

Therefore rule 30 (and for that matter rule 110, game of life ...)
shall be analysed as a composite function with exponential description
complexity growth w.t.r. to the input growth. 

From there it is easy to argue irreducibility of rule 30 because of
basic property of composite function, the output of one function determines
input to other. Any reduction in function composition will mean reduction
of branching structure. In case of rule 30, the output of the previous
function determines the input \emph{and selection} of the next function
in the composite function chain. That detail prevents any short-cut
execution path through algorithm.

\section{Conclusion\label{sec:Conclusion}}

\subsection{Summary\label{sub:Summary} }

As is noted in the introduction, the whole discussion is about three
notions:
\begin{enumerate}
\item Structured Programming Theorem; particularly treating branching as
basic structure in programming. 
\item Cyclomatic Complexity; exponential dependence between branching and
number of execution paths a program can take. Basically every branching
potentially doubles number of paths. 
\item Formal description of an algorithm; requirement that every branching
in algorithm shall be fully defined.
\end{enumerate}
The first option is: one or more of above do not hold. The second
option is: all above notions hold and there exists a program without
any knowledge of output behaviour before input is presented. The discussion
from this paper sees the second option as a true. The main arguments
for this are:
\begin{itemize}
\item The 3n+1 parity sequence can be used as encoding system, see subsection
\ref{sub:3n+1-as-encoding system}. The argument is that 3n+1 encoding
alias function description can not be simpler than standard binary
encoding of an input. 
\item The 3n+1 algorithm description exponential growth can be reduced only
if branching structure can be reduced to sequential and iteration
programming structure in polynomial time. See subsections \ref{sub:3n+1-and-reductions}
and \ref{sub:Rule-30-and irreducibility}.
\item The random oracle framework provides the definition of correlation
intractability and how that requirement can not be obtained by single
function or function assembly (see quote subsection \ref{sub:3n+1-as-a oracle}).
Contrary to that notion 3n+1 algorithm looks like:$(\textipa{ select \ensuremath{f}or \ensuremath{g)\circ(\textipa{ select \ensuremath{f}or \ensuremath{g)\circ}}(\textipa{ select \ensuremath{f}or \ensuremath{g)\ldots}}}}$.
It is apparent that function description without specific input is
not present, and that the input actually defines function composition.
Therefore there is the case when input description and function description
have the same complexity. That case satisfies the correlation intractability
requirement (subsection \ref{sub:3n+1-as-a oracle}). 
\end{itemize}
Other arguments are various empirical findings, for example rule 30
is used as random number generator in Wolfram's Mathematica.

The common features of 3n+1 and rule 30 are:
\begin{itemize}
\item Composed from two distinctive functions $f$ and $g$ that are not
commutative $f\circ g\neq g\circ f$.
\item Cyclomatic Complexity raise with every branching step see subsection
\ref{sub:Rule-30-function description}
\item Steps in program execution path are one of the function $f$ or $g$ 
\item Probability of executing $f$ or $g$ in next step is $0.5$
\end{itemize}
See Appendix A for another practical implementation of above features.

\subsection{Randomness and simple arithmetic?}

As is discussed in subsection \ref{sub:3n+1-as-a oracle} the proposition
is to exchange tables \ref{tab:Random-Oracle-mapping} and \ref{tab:3n+1-parity-mapping}
without loosing any of the random oracle properties. The similarity
between tables is: Both tables are impractical if used in tabular
form. There is a storage problem (for example how to store $2^{128}$entries
? and seek time cost). It is easy to see that random oracle table
can not be compressed because the second column is by design true
random. On the opposite side of tabular representation spectrum is
binary encoding (table \ref{tab:binary-encoding}). If input is given
in the left column of the binary table it is easy to calculate corresponding
entry in the right column and vice versa. This means that the tabular
form is not needed (easily calculated/compressed) because it is easy
to calculate entries both ways. 

\begin{table}[h]
\begin{tabular}{|c|c|}
\hline 
$n$ & binary encoding\tabularnewline
\hline 
\hline 
$1$ & $1$\tabularnewline
\hline 
$2$ & $10$\tabularnewline
\hline 
$3$ & $11$\tabularnewline
\hline 
$4$ & $100$\tabularnewline
\hline 
$\ldots$ & $\ldots$\tabularnewline
\hline 
\end{tabular}

\caption{binary encoding\label{tab:binary-encoding}}

\end{table}

Because 3n+1 can be used as encoding system (subsection \ref{sub:3n+1-as-encoding system}),
the compression of table \ref{tab:3n+1-parity-mapping} is not an
issue. 

Now it is time to see how the proposition from subsection \ref{sub:3n+1-as-a oracle}
reflects on parity and binary encoding table:
\begin{itemize}
\item Binary table is not affected if only part of the string in the right
hand column is provided. For example, if the question is to find corresponding
entries for binary string beginning with $10$, just appending arbitrary
suffix to $10$ and decoding that string will find entry on the left
hand side column..
\item The parity table in the case when only a partial string in the right
hand column is provided can not be calculated or compressed. The simple
reason is that entries can be calculated only with complete input.
Anything else faces the ambiguous prospect of $(f\clubsuit g)\circ(f\clubsuit g)\circ(f\clubsuit g)\ldots$
where $f\clubsuit g$ means depending on input use function $f$ or
use function $g$.
\end{itemize}
The question is: can observation of entries in parity table (table
\ref{tab:3n+1-parity-mapping}) provide any means of compressing that
table? The answer is no, because the branching structure of algorithm
prevents any type of Solomon-Kolgomorov-Chaitin (SKC) reductions,
even though the table is deterministic in nature. In other words the
data in the parity table ought to be random. Rule 30 sequences are
in the same category. It is remarkable that randomness can be now
interpreted as inability of reducing selection criteria programming
structure. Translated to random oracle vocabulary that is notion of
correlation intractability.

\subsection{P and NP}

The 3n+1 proposal for collision resistance (subsection \ref{sub:3n+1-as-a oracle})
can serve as P versus NP discussion as well. The game is to find input
$x$ (natural number) and with that $x$ to produce the parity string
$p_{x}$. Parity $p_{x}=s||a$ is the concatenation of given string
$s$ and arbitrary string $a$. Only one constraint is $l_{x}=2l_{s}$,
where $l_{x}$ is the binary length of $x$ and $l_{s}$ is the binary
length of the given string $s$. 

For example the given string in C language notation is \emph{char
}$s="DoesPequalsNp?"$ has the binary length $l_{x}=14*8$. The task
is to find natural number $x$ with binary length $l_{x}=2*14*8$
and with sequence \emph{char} $p_{x}="DoesPequalsNp?...$''. 

First of all, nothing guaranties that any of the natural numbers 224
bit long will produce required parity sequence.

Secondly, because matching parity is not fully defined calculating
$x$ from $s$ is impossible. The reasons are: 
\begin{itemize}
\item To compose the transformation and do the calculation full knowledge
of input is needed, because only input defines function composition.
\item Trying to observe the mapping of natural numbers to corresponding
parities and hoping to find some pattern/reduction is futile because
the selection criteria programming structure can not be reduced.
\end{itemize}
\begin{table}[h]
\begin{tabular}{|c|}
\hline 
$(f\clubsuit g)\circ(f\clubsuit g)\circ(f\clubsuit g)$\tabularnewline
\hline 
\hline 
$f\circ f\circ f$\tabularnewline
\hline 
$f\circ f\circ g$\tabularnewline
\hline 
$f\circ g\circ f$\tabularnewline
\hline 
$f\circ g\circ g$\tabularnewline
\hline 
$g\circ f\circ f$\tabularnewline
\hline 
$g\circ f\circ g$\tabularnewline
\hline 
$g\circ g\circ f$\tabularnewline
\hline 
$g\circ g\circ g$\tabularnewline
\hline 
\end{tabular}

\caption{The complexity of 3n+1 function composition grows exponentially; example
of function 3 steps possible outcomes\label{tab:Composition-function-outcome}}

\end{table}

The game has all the main ingredients of $P\neq NP$. If complete
input (either natural number or parity) is provided it is easy to
do 3n+1 encoding (subsection \ref{sub:3n+1-as-encoding system}).
If input is not fully defined the only technique for finding natural
number/parity pair is exhaustive search. That exhaustive search is
exponential in nature (table \ref{tab:Composition-function-outcome}).
On the other hand if match is found it is easy to verify that because
complete input is now known. 

\appendix

\section{Sharatz.h listing\label{sec:Sharatz.h-listing}}

Here is an application of the principles argued in the paper. Using
features mentioned in subsection \ref{sub:Summary} the program outputs
a random stream. The program is a variant of MAG 2. The latest known
cryptanalysis of MAG family with all details is presented here \cite{key-28}
including review of published pseudo attack.

The function accepts seed and desired length of the stream. The produced
stream is directed to standard output and can be piped or saved to
the file.

\begin{lstlisting}[caption={sharatz.h},captionpos={b  //--> caption at the bottom of the listing (default is "t")},label={sarac}]
#include <stdio.h>
#include <stdint.h>
#include <string.h>
#include <stdlib.h>

typedef enum { OK = 0, FAIL = 1} sarac_return;

typedef char seed_sequence;//max 8k

sarac_return sarac 
(const seed_sequence * sarac_seed, uint64_t sarac_byte_len)
{
	int i, j, seed_len;
	int rounds = 4;

	uint64_t mixer = 6148914691236517205;
	uint64_t ignite[1026];
	uint64_t carry = 1234567890123456789;
	uint64_t ying[512], yang[512];
	uint64_t carry_a, carry_b;
	uint64_t data_block, k;
	uint64_t output;
    
	for(i = 0; i < 1026; i++)
		ignite[i] = 0;

	seed_len = strlen(sarac_seed);
    
	if(seed_len > 8 * 1026)
		seed_len = 8 * 1026;

	memcpy(ignite, sarac_seed, seed_len);

	for(i = 0; i < rounds; i++)
	{
		for(j = 0; j < 1026; j++)
		{
			if(ignite[(j+2)%1026]>ignite[(j+3)%1026])
				carry ^= ignite[(j+1)%1026];
			else
				carry ^= ~ignite[(j+1)%1026];

			ignite[j] ^= carry;
			carry += mixer;
		}
	}
   
	for(i = 0; i < 512; i++)
		ying[i] = ignite[i];
		
	for(i = 0; i < 512; i++)
		yang[i] = ignite[i+512];

	carry_a = ignite[1024];
	carry_b = ignite[1025];
	
	data_block = sarac_byte_len / 8 ;
	if((sarac_byte_len % 8) > 0)
		data_block++;

	for(k = 0; k < data_block; k++)
	{
		if(ying[(k+2)%512]>ying[(k+3)%512])
			carry_a ^= ying[(k+1)%512];
		else
			carry_a ^= ~ying[(k+1)%512];
				
		ying[k%512] ^= carry_a;
			
		if(yang[(k+2)%512]>yang[(k+3)%512])
			carry_b ^= yang[(k+1)%512];
		else
			carry_b ^= ~yang[(k+1)%512];
				
		yang[k%512] ^= carry_b;
			
		output = ying[k%512] ^ yang[k%512];
					
		fwrite(&output, sizeof(output),1,stdout);
	}

	return OK;

}
\end{lstlisting}
\end{document}